\documentclass[12pt,a4paper]{article}
\pdfoutput=1
\usepackage{graphicx}
\usepackage[T1]{fontenc}
\usepackage[sc,osf]{mathpazo}
\usepackage{a4wide}  
\usepackage{latexsym,amsthm,amsfonts,amsmath,mathrsfs,amssymb}
\usepackage[unicode,implicit]{hyperref}
\hypersetup{%
  pdftitle    = {On gauged maximal d=8 supergravities}
  pdfkeywords = {gravity, supergravity, gauge symmetry, Yang-Mills, embedding tensor, tensor hierarchy},
  pdfauthor   = {Oscar Lasso Andino and Tomas Ortin},
  plainpages  = true,
  colorlinks  = true,
  citecolor   = blue,
  urlcolor    = red,
  linkcolor   = black
}
\newcommand{\hepth}[1]{{\tt
\href{http://www.arXiv.org/abs/hep-th/#1}{hep-th/#1}}}

\newcommand{\arxiv}[1]{{\tt
\href{http://www.arXiv.org/abs/#1}{#1}}}
\newcommand{\doi}[1]{{\tt DOI:\href{http://dx.doi.org/#1}{#1}}}
\makeatletter
\@addtoreset{equation}{section}
\makeatother

\begin{document}

\begin{flushright}
\small
IFT-UAM/CSIC-16-046\\
\texttt{arXiv:1605.09629 [hep-th]}\\
December 14\textsuperscript{th}, 2017\\
\normalsize
\end{flushright}

\vspace{1.5cm}

\begin{center}

{\Large {\bf On gauged maximal $d=8$ supergravities}}

\vspace{2.5cm}

\renewcommand{\thefootnote}{\alph{footnote}}
{\sl\large  \'Oscar Lasso Andino}\footnote{E-mail: {\tt oscar.lasso [at] estudiante.uam.es}}
{\sl\large and Tom\'{a}s Ort\'{\i}n}\footnote{E-mail: {\tt Tomas.Ortin [at] csic.es}}

\setcounter{footnote}{0}
\renewcommand{\thefootnote}{\arabic{footnote}}

\vspace{1.5cm}

{\it Instituto de F\'{\i}sica Te\'orica UAM/CSIC\\
C/ Nicol\'as Cabrera, 13--15,  C.U.~Cantoblanco, E-28049 Madrid, Spain}\\ \vspace{0.3cm}

\vspace{2.8cm}


{\bf Abstract}

\end{center}

\begin{quotation}
  We study the gauging of maximal $d=8$ supergravity using the embedding
  tensor formalism. We focus on SO$(3)$ gaugings, study all the possible
  choices of gauge fields and construct explicitly the bosonic actions
  (including the complicated Chern-Simons terms) for all these choices, which
  are parametrized by a parameter associated to the 8-dimensional
  SL$(2,\mathbb{R})$ duality group that relates all the possible choices which
  are, ultimately, equivalent from the purely 8-dimensional point of view.

  Our result proves that the theory constructed by Salam and Sezgin by
  Scherk-Schwarz compactification of $d=11$ supergravity and the theory
  constructed in Ref.~\cite{AlonsoAlberca:2000gh} by dimensional reduction of
  the so called ``massive 11-dimensional supergravity'' proposed by Meessen
  and Ort\'{\i}n in Ref.~\cite{Meessen:1998qm} are indeed related by an
  SL$(2,\mathbb{R})$ duality even though they have two completely different
  11-dimensional origins.

\end{quotation}

\newpage
\pagestyle{plain}

\tableofcontents


\section{Introduction}

Gauged/massive supergravities have received a great deal of attention over the
last few years because they almost always include a scalar potential that
could fix the moduli or provide an interesting inflationary model. While
gauging a given supergravity theory obtained, for instance, as the low-energy
limit of some string theory model, is just a technical problem which we now
know how to handle in general, the string-theory description of that gauged
theory, its 11-dimensional origin and the meaning of the new constants that
appear in it (coupling constants, mass parameters etc.), are not always known.

The gauging of maximal 8-dimensional supergravity offers a particularly
interesting example. Salam and Sezgin obtained this theory with an
$\mathrm{SO}(3)\subset \mathrm{SL}(3,\mathbb{R})$ gauging in
Ref.~\cite{Salam:1984ft} by performing a Scherk-Schwarz reduction
\cite{Scherk:1979zr} of 11-dimensional supergravity
\cite{Cremmer:1978km}.\footnote{Other 3-dimensional groups can be obtained by
  the same procedure, as shown in
  Refs.~\cite{AlonsoAlberca:2003jq,Bergshoeff:2003ri}. We also remind the
  reader of the U-duality group of this theory: it is
  SL$(2,\mathbb{R})\times \mathrm{SL}(3,\mathbb{R})$} The gauge
fields of this theory are the three Kaluza-Klein vectors. However, the theory
has another SL$(3,\mathbb{R})$ triplet of vectors that can be used as
gauge fields: the vectors that come from the 11-dimensional 3-form. This
alternative $\mathrm{SO}(3)$ gauging can be carried out directly in 8
dimensions by the standard methods, but it is not known how to obtain this
theory from the conventional 11-dimensional supergravity.

Actually, it is believed that it should be possible to obtain this second
SO$(3)$-gauged theory by an SL$(2)$ rotation of the Salam-Sezgin one. These
transformations of gauged theories are no longer symmetries of their equations
of motion. Rather, they are (very complicated) field redefinitions. Thus, at a
classical level, and from the 8-dimensional point of view, these two theories
should be equivalent.

From the 11-dimensional point of view, the situation is less clear: on the one
hand,in principle one may use the 8-dimensional relation between the fields in
the two theories to construct a very unnatural and non-local\footnote{The
  SL$(2)$ transformation that should relate these two SO$(3)$ gauged theories
  involves electric-magnetic rotations of the 3-form potential.} alternative
compactification Ansatz which would give the second SO$(3)$ gauged theory
instead of the Salam-Sezgin one. On the other hand, it is hard to say whether
these two theories are equivalent from the 11-dimensional point of view.

It is somewhat surprising that the second SO$(3)$-gauged maximal supergravity
can be obtained with \textit{exactly the same compactification Ansatz} as the
Salam-Sezgin one from the so-called \textit{massive 11-dimensional
  supergravity} \cite{AlonsoAlberca:2000gh}. This theory is a deformation of
11-dimensional supergravity proposed in Ref.~\cite{Meessen:1998qm} as a
candidate to 11-dimensional origin of Romans' massive $\mathcal{N}=2A,d=10$
supergravity \cite{Romans:1985tz}.\footnote{The supersymmetry transformations
  of this theory were studied in Ref.~\cite{Gheerardyn:2001jj}.} This theory
does not have 11-dimensional covariance, as it depends explicitly on the
(commuting) Killing vectors but, somewhat mysteriously, it turns out that it
can account for the 11-dimensional origin of several gauged supergravity
theories (apart from Romans' and the 8-dimensional one under discussion) which
are not obtained, wit the conventional compactification Ansatz, by standard
methods \cite{AlonsoAlberca:2002tb}.

Our goal in this paper twofold: first, we want to show that the gauged theory
obtained from the compactification of \textit{massive 11-dimensional
  supergravity} (which will be referred to henceforth as AAMO) is indeed one
of the SO$(3)$-gauged maximal supergravities that can be obtained using the
embedding tensor method. Second, we want to show that, from the 8-dimensional
point of view, it is related to the Salam-Sezgin one (from this moment, SS) by
an SL$(2,\mathbb{R})$ transformation. We will achieve both goals by
constructing a 1-parameter family of SL$(2)$-related SO$(3)$-gauged
supergravities that interpolates between the SS and AAMO
theories.\footnote{The existence of these duality-related family of gaugings
  has been noticed in Refs.~\cite{deRoo:2011fa,Dibitetto:2012rk}.}

The best way to construct these gauged theories is through the use of the
embedding tensor formalism
\cite{Cordaro:1998tx,deWit:2002vt,deWit:2003hq,deWit:2005hv,deWit:2005ub}.\footnote{For
  recent reviews see
  Refs.~\cite{Trigiante:2007ki,Weidner:2006rp,Samtleben:2008pe,Ortin:2015hya}.}
This formalism has been used in several maximal and half-maximal
supergravities
Refs.~\cite{deWit:2004nw,Samtleben:2005bp,Schon:2006kz,deWit:2007kvg,Bergshoeff:2007vb,deWit:2008ta,FernandezMelgarejo:2011wx}. In
the 8-dimensional case it has been used in Ref.~\cite{Puigdomenech:2008kia} to
study the possible subgroups of the U-duality group that can be gauged,
regardless of the vectors used as gauge fields, by solving the constraints
satisfied by the embedding tensor. The existence of continuous families of
gauged supergravities escapes this kind of analysis, though, and we are
actually interested in the explicit construction of the theory. In a more
recent paper \cite{Andino:2016bwk} we have used the embedding-tensor formalism
to construct the most general 8-dimensional gauge theory (including its tensor
hierarchy), for any field content and duality group. This result can
immediately be particularized to the field content, $d$-tensors and duality
group of the maximal 8-dimensional supergravity and we just have to find a
1-parameter SO$(3)$ solution of the constraints satisfied by the embedding
tensor an other deformation parameters to have the complete tensor hierarchy
of the theory we are after. To end the construction of the bosonic theory it
only remains to find the scalar potential and the equations of motion. We will
explain how to do that in this case. We will also explain how to construct the
supersymmetry transformation rules.

This paper is organized as follows: in Section~\ref{sec-ungauged} we review
the matter content and symmetries of the ungauged theory. We will introduce a
new basis of fields with simpler transformation properties, as required by the
embedding tensor formalism. In Section~\ref{sec-gauging} we will discuss the
gauging, using that formalism, of the theory, applying the general results of
Ref.~\cite{Andino:2016bwk}. We will show that there is a 1-parameter family of
embedding and other deformation tensors associated to SO$(3)$ gaugings we are
after. In Section~\ref{sec-construction} we proceed to the explicit
construction of the theory. Our conclusions are described in
Section~\ref{sec-conclusions} and, in the appendices, the explicit forms of
the field strengths, Bianchi identities, identities of Bianchi identities and
duality relations, are collected.

\section{Ungauged $\mathcal{N}=2$, $d=8$ Supergravity}
\label{sec-ungauged}

In this section we are just going to describe the aspects of the ungauged
theory that we need to know in order to construct the family of gauged
supergravities we are after.

$\mathcal{N}=2$, $d=8$ supergravity can be obtained by direct dimensional
reduction of 11-dimensional supergravity on $T^{3}$ \cite{Salam:1984ft}. The
scalars of the theory parametrize the coset spaces
SL$(2,\mathbb{R})/\mathrm{SO}(2)$ and
SL$(3,\mathbb{R})/\mathrm{SO}(3)$. The U-duality group of the theory
is SL$(2,\mathbb{R})\times\mathrm{SL}(3,\mathbb{R})$ and its fields
are either invariant or transform in the fundamental representations of both
groups. We use the indices $i,j,k=1,2$ for SL$(2,\mathbb{R})$
doublets and $m,n,p=1,2,3$ for SL$(3,\mathbb{R})$ triplets.

The bosonic fields are

\begin{equation}
g_{\mu\nu}, C, B_{m}, A^{i\, m}, a,\varphi,\mathcal{M}_{mn},
\end{equation}

\noindent
where $C$ is a 3-form, $B_{m}$ a triplet of 2-forms, $A^{i\, m}$, a doublet of
triplets of 1-forms (six in total), $a$ and $\varphi$ are the axion and
dilaton fields which can be combined into the axidilaton field

\begin{equation}
\tau\equiv a+i e^{-\varphi}\, ,  
\end{equation}

\noindent
or into the SL$(2,\mathbb{R})/\mathrm{SO}(2)$ symmetric matrix

\begin{equation}
\label{eq:Wijdef}
\left(\mathcal{W}_{ij}\right)
\equiv
e^{\varphi}
\left(
  \begin{array}{cc}
  |\tau|^{2} & a \\
a & 1 \\ 
  \end{array}
\right)\, ,  
\,\,\,\,\,
\mbox{with inverse}
\,\,\,\,\,
\left(\mathcal{W}^{ij}\right)
\equiv
e^{\varphi}
\left(
  \begin{array}{cc}
  1 & -a \\
-a & |\tau|^{2} \\ 
  \end{array}
\right)\, ,  
\end{equation}

\noindent
and, finally, $\mathcal{M}_{mn}$ is an
SL$(3,\mathbb{R})/\mathrm{SO}(3)$ symmetric matrix whose explicit
parametrization in terms of five independent scalars will not concern us for
the moment. The inverses of these matrices will be written with upper indices.

The bosonic action obtained in Ref.~\cite{AlonsoAlberca:2000gh} by simple
dimensional reduction is\footnote{ The relation between the 8- and
  11-dimensional fields can be found there. As mentioned in
  Ref.~\cite{AlonsoAlberca:2000gh}, one of the coefficients in the
  Chern-Simons part of the action (which has been checked explicitly to be
  gauge-invariant) differs from the corresponding one in
  Ref.~\cite{Salam:1984ft}.}

\begin{equation}
\label{eq:d8ungaugedaction}
\begin{array}{rcl}
S 
& = & 
{\displaystyle\int} d^{8}x \sqrt{|g|}\,
\left\{ 
R
+\frac{1}{4}{\rm Tr}\left(\partial \mathcal{M}\mathcal{M}^{-1}\right)^{2}
+\frac{1}{4}{\rm Tr}\left(\partial \mathcal{W}\mathcal{W}^{-1}\right)^{2}
\right.
\\
& & \\
& & 
-\frac{1}{4}F^{i\, m}\mathcal{M}_{mn}\mathcal{W}_{ij} F^{j\, n}
+\frac{1}{2\cdot 3!} H_{m}\mathcal{M}^{mn} H_{n}
-\frac{1}{2\cdot 4!} e^{-\varphi} G^{2} \, ,
\\
& & \\
& & 
-\frac{1}{6^{3}\cdot 2^{4}}
{\textstyle\frac{1}{\sqrt{|g|}}}\, \epsilon
\left[GGa -8G H_{m} A^{2\, m} +12 G(F^{2\, m}+aF^{1\, m})B_{m} 
\right.
\\
&& \\
& & 
\left.\left.
-8\epsilon^{mnp} H_{m}H_{n}B_{p} -8G\partial a C 
-16 H_{m}(F^{2\, m}+aF^{1\, m}) C \right]\right\} \, ,
\\
\end{array}
\end{equation}

\noindent 
where the field strengths are given by\footnote{In this notation, used in
  Ref.~\cite{AlonsoAlberca:2000gh}, all the lower indices, which are not
  shown, are antisymmetrized with weight one. The difference with
  differential-form notation is the normalization of the components of the
  differential forms: $\omega^{(p)}=\tfrac{1}{p!}\omega^{(p)}_{\mu_{1}\cdots
    \mu_{p}}dx^{\mu_{1}}\wedge \cdots dx^{\mu_{p}}$, so, for instance,
  $d\omega^{(p)}= (p+1)\partial \omega^{(p)}$.}

\begin{equation}
\label{eq:fieldstrengths1}
\begin{array}{rcl}
F^{i\, m} 
& = &
2\partial A^{i\, m}\, .
\\
& & \\
H_{m} 
& = &
3\partial B_{m} +3\epsilon_{mnp}F^{1\, n}A^{2\, p}\, ,
\\
& & \\
G 
& = &
4 \partial C +6 F^{1\, m} B_{m}\, ,
\\
\end{array}
\end{equation}

\subsection{Rewriting the theory}
\label{sec-rewritting}

In order to study the gaugings of this theory using the embedding-tensor
formalism it is convenient to use differential-form language and, furthermore,
use a different basis of forms with better transformation properties under the
duality groups (in particular, under SL$(2,\mathbb{R})$): for
instance, if the 3-form field strengths $H_{m}$ are invariant under
SL$(2,\mathbb{R})$ transformations, it is obvious that the 2-forms
$B_{m}$ can only be invariant under those SL$(2,\mathbb{R})$
transformations up to 1-form gauge transformations because the Chern-Simons
term $3\epsilon_{mnp}F^{1n}A^{2p}$ has that same behaviour. This implies, in
its turn, that the 3-form $C$ only transforms as the first component of an
SL$(2,\mathbb{R})$ doublet (something we expect to happen on
general grounds) up to gauge transformations. The conclusion is that we are going
to need to redefine the 2- and 3-form potentials $B_{m}$ and $C$, which we
also denote as $C^{1}$ when needed.

In differential-form language, the above field strengths take the form

\begin{equation}
\label{eq:fieldstrengths2}
\begin{array}{rcl}
F^{im} 
& = &
d A^{im}\, .
\\
& & \\
H_{m} 
& = &
d B_{m} +\epsilon_{mnp}F^{1n}\wedge A^{2p}\, ,
\\
& & \\
G 
& = &
dC + F^{1m}\wedge B_{m}\, .
\\
\end{array}
\end{equation}

\noindent
The redefinition of the potentials that gives the the required properties of
transformation under the U-duality group is

\begin{equation}
  \begin{array}{rcl}
B_{m} 
& \longrightarrow &
B_{m} -\tfrac{1}{2}\epsilon_{mnp}A^{1n}\wedge A^{2p}\, , 
\\
& & \\
C
& \longrightarrow &
C^{1} +\tfrac{1}{2}\epsilon_{mnp}A^{1m}\wedge A^{1n}\wedge A^{2p} \, .
\end{array}
\end{equation}

\noindent
In terms of these new potentials, the field strengths take the
form\footnote{Here and, very often in what follows, we suppress the wedge
  product symbols $\wedge$ in order to simplify the expressions. We will
  introduce further simplifications in the notation along the way.}

\begin{eqnarray}
\label{eq:Fimungauged}
F^{im} 
& = &
d A^{im}\, ,
\\
& & \nonumber \\
\label{eq:Hmungauged}
H_{m} 
& = &
d B_{m} +\tfrac{1}{2}\epsilon_{ij}\epsilon_{mnp}F^{in}A^{jp}\, ,
\\
& & \nonumber \\
G^{1} 
& = &
dC^{1} + F^{1m}B_{m}
+\tfrac{1}{6}\epsilon_{ij}\epsilon_{mnp}A^{1m}F^{in}A^{jp}\, .
\end{eqnarray}

\noindent
The gauge transformations that leave these field strengths invariant are

\begin{equation}
\label{eq:gaugeungauged1}
\begin{array}{rcl}
\delta_{\sigma}A^{im} 
& = &
d \sigma^{im}\, .
\\
& & \\
\delta_{\sigma} B_{m} 
& = &
d \sigma_{m} -\epsilon_{ij}\epsilon_{mnp}
\left( F^{in}\sigma^{jp}-\tfrac{1}{2}A^{in}\delta_{\sigma}A^{jp} \right)\, ,
\\
& &  \\
\delta_{\sigma}C^{1} 
& = &
d\sigma^{1} 
-\left[
\sigma^{1m}H_{m} +F^{1m}\sigma_{m} +\delta_{\sigma}A^{1m}B_{m}
+\tfrac{1}{6}\epsilon_{jk}\epsilon_{mnp}\delta_{\sigma}A^{jn}A^{1m}A^{kp}
\right]\, ,
\end{array}
\end{equation}

\noindent
and the gauge-invariant bosonic action can be written in the form

\begin{equation}
\label{eq:d8ungaugedactiondiffform}
\begin{array}{rcl}
S 
& = & 
{\displaystyle\int} 
\left\{ 
-\star R
+\frac{1}{4}{\rm Tr}
\left(
d \mathcal{M}\mathcal{M}^{-1}
\wedge 
\star d \mathcal{M}\mathcal{M}^{-1} \right)
+\frac{1}{4}{\rm Tr}
\left(
d \mathcal{W}\mathcal{W}^{-1}
\wedge 
\star d \mathcal{W}\mathcal{W}^{-1} \right)
\right.
\\
& & \\
& & 
+\frac{1}{2}\mathcal{W}_{ij}\mathcal{M}_{mn}F^{im}\wedge \star F^{jn}
+\frac{1}{2} \mathcal{M}^{mn} H_{m}\wedge \star H_{n}
+\frac{1}{2} e^{-\varphi} G^{1} \wedge \star G^{1}
-\frac{1}{2} a G^{1}  G^{1}
\\
& & \\
& & 
+\frac{1}{3} G^{1} 
\left[
H_{m} A^{2m} -B_{m} F^{2m} 
+\frac{1}{2}\epsilon_{mnp}F^{2m} A^{1n} A^{2p}
\right]
\\
& & \\
& & 
+\frac{1}{3} H_{m} F^{2m} 
\left[
C^{1} 
+
\frac{1}{6}\epsilon_{mnp}A^{1m} A^{1n} A^{2p}
\right]
\\
&& \\
& & 
\left.
+\frac{1}{3!}\epsilon^{mnp} H_{m} H_{n} 
\left(
B_{p} -\tfrac{1}{2}\epsilon_{pqr}A^{1q} A^{2r}
\right)
\right\} \, .
\\
\end{array}
\end{equation}

It is not difficult to check that the (formal\footnote{It is the total
  derivative of an 8-form in 8 dimensions.}) exterior derivative of the
Chern-Simons part of this action (the last three lines) is just a combination
of gauge-invariant field strengths:

\begin{equation}
d(\mathrm{Chern-Simons})
=
-H_{m} F^{2m} G^{1} -\tfrac{1}{3!}\epsilon^{mnp} H_{m} H_{n}
 H_{p}\, ,  
\end{equation}

\noindent
which ensures its gauge-invariance up to total derivatives under the
transformations Eqs.~(\ref{eq:gaugeungauged1}).

Global SL$(2,\mathbb{R})$ covariance requires the introduction of
another 3-form $C^{2}$ so we can define a doublet of 4-form field strengths

\begin{equation}
\label{eq:Giungauged}
G^{i} 
\equiv
dC^{i} + F^{im}B_{m}
+\tfrac{1}{6}\epsilon_{jk}\epsilon_{mnp}A^{im}F^{jn}A^{kp}\, ,
\end{equation}

\noindent
invariant under the gauge transformations $\delta_{\sigma}A^{im}$ and
$\delta_{\sigma}B_{m}$ in Eq.~(\ref{eq:gaugeungauged1}) and

\begin{equation}
\label{eq:gaugeungauged2}
\delta_{\sigma}C^{i} 
= 
d\sigma^{i} 
-\left[
\sigma^{im}H_{m} +F^{im}\sigma_{m} +\delta_{\sigma}A^{im}B_{m}
+\tfrac{1}{6}\epsilon_{jk}\epsilon_{mnp}\delta_{\sigma}A^{jn}A^{im}A^{kp}
\right]\, .  
\end{equation}

This (magnetic, dual) field is related by electric-magnetic duality to the
original (electric, fundamental) $C$ so there are no new degrees of freedom
on \textit{duality shell}\footnote{Observe that $ \tilde{G}$ is a combination
  of the field strength of the electric 3-form $G$, its Hodge dual $\star G$
  and the scalars, while $G^{2}$ is the field strength of the magnetic
  3-form $C^{2}$. }

\begin{equation}
G^{2} = e^{-\varphi} \star G +a G \equiv \tilde{G}\, ,
\end{equation}

\noindent
and the relation is such that, using it, the equation of motion of $C$ that
follows from the action Eq.~(\ref{eq:d8ungaugedactiondiffform})

\begin{equation}
-\frac{\delta S}{\delta C}
= 
d \tilde{G}-F^{1m} H_{m}\, ,   
\end{equation}

\noindent
becomes the Bianchi identity for the field strength $G^{2}$. 

Then, denoting with a $\Delta$ the part of a $(p+1)$-field strength that does
not contain the derivative of the $p$-form potential, using the above
definitions we can rewrite the action Eq.~(\ref{eq:d8ungaugedactiondiffform})
in a more compact form that we will use later:

\begin{equation}
\label{eq:d8ungaugedactiondiffform2}
\begin{array}{rcl}
S 
& = & 
{\displaystyle\int} 
\left\{ 
-\star  R
+\frac{1}{4}{\rm Tr}
\left(
d \mathcal{M}\mathcal{M}^{-1}
\wedge 
\star d \mathcal{M}\mathcal{M}^{-1} \right)
+\frac{1}{4}{\rm Tr}
\left(
d \mathcal{W}\mathcal{W}^{-1}
\wedge 
\star d \mathcal{W}\mathcal{W}^{-1} \right)
\right.
\\
& & \\
& & 
+\frac{1}{2}\mathcal{W}_{ij}\mathcal{M}_{mn}F^{im}\wedge \star F^{jn}
+\frac{1}{2} \mathcal{M}^{mn} H_{m}\wedge \star H_{n}
+\frac{1}{2} G  \tilde{G}
-dC^{1}\Delta G^{2} -\tfrac{1}{2}\Delta G^{1}\Delta G^{2} 
\\
& & \\
& & 
\left.
-\tfrac{1}{12}\epsilon^{mnp}B_{m}dB_{n}dB_{p}
+\tfrac{1}{4}\epsilon^{mnp}B_{m}H_{n}H_{p}
-\tfrac{1}{24}\epsilon_{ij} A^{im} A^{jn} \Delta H_{m}dB_{n}
\right\} \, .
\\
\end{array}
\end{equation}

Potentials dual to the 2-forms (the 4-forms $\tilde{B}^{m}$), to the 1-forms
$A^{im}$ (the 5-forms $\tilde{A}_{im}$) and to the scalars (the 6-forms
$D_{A}$, where the index $A$ runs over the adjoint representation of the
duality group SL$(2,\mathbb{R})\times \mathrm{SL}(3,\mathbb{R})$),
and their gauge-invariant field strengths
($\tilde{H}^{m},\tilde{F}_{im},K_{A}$) can also be defined by dualizing the
equations of motion of the corresponding electric fields. We will not need
them now, but they can be found in Ref.~\cite{Andino:2016bwk}. They can also be
recovered by setting to zero the deformation parameters in the field strengths
of the gauged theory that we are going to construct in the next section and
which are listed in Appendix~\ref{sec-fieldstrengths}.

\section{SO$(3)$ gaugings of $\mathcal{N}=2$, $d=8$ supergravity}
\label{sec-gauging}

The gaugings and massive deformations of general 8-dimensional field theories
have been studied in depth using the embedding-tensor formalism in
Ref.~\cite{Andino:2016bwk} using the notation of Ref.~\cite{Ortin:2015hya} and
the general procedure used in Refs.~\cite{Bergshoeff:2009ph,Hartong:2009vc}
for the 4-,5- and 6-dimensional cases: finding identities for Bianchi
identities, solving those identities for the Bianchi identities and then
solving the Bianchi identities for the field strengths. In particular, the
tensor hierarchy has been constructed and the form of most of the field
strengths has been fully determined. The action was only determined up to
terms containing 2-forms due to the very large number of complicated terms
occurring in it.

In this section we are going to specialize the results of
Ref.~\cite{Andino:2016bwk} to the particular case of $\mathcal{N}=2$, $d=8$
supergravity and, then, we are going to select the family of gaugings we are
interested in\footnote{A partial analysis of the possible gaugings (that is:
  the possible solutions to the constraints satisfied by the embedding tensor)
  was performed in Ref.~\cite{Puigdomenech:2008kia}.}. Since the case we are
going to study is far simpler than the general case, we are going to determine
almost the bosonic action.

In order to particularize the results of Ref.~\cite{Andino:2016bwk} to
$\mathcal{N}=2$, $d=8$ supergravity we have to particularize the generic field
content, the $d$-tensors occurring in the Chern-Simons terms and the global
symmetry group considered there.

Let us start by reviewing the U-duality group of the theory. The U-duality
group of this theory is, exactly,
SL$(2,\mathbb{R})\times$SL$(3,\mathbb{R})$\footnote{There is only one
  additional rescaling symmetry, but it acts on the spacetime metric and,
  therefore, we will not consider it here.}  and we remind the reader of the
group isomorphism SL$(2,\mathbb{R})\sim\mathrm{Sp}(2,\mathbb{R})$.  The
adjoint indices of the U-duality group are denoted collectively by
$A,B,\ldots$. The adjoint indices of SL$(2,\mathbb{R})$ are
$\alpha,\beta,\ldots =1,2,3$. The adjoint indices of SL$(3,\mathbb{R})$ are
$m,n,\ldots=1,2,3$ for the SO$(3)$ subgroup that we want to gauge and
$a,b,\ldots=1,\cdots,5$ for the rest of the generators.

The only structure constants that we need to know explicitly are those of the
SO$(3)$ subgroup:\footnote{SO$(3)$ indices are raised and lowered with the
  unit metric and, therefore, there is no distinction between upper and lower
  SO$(3)$ indices. We choose their position for the sake of convenience and
  esthetics.}

\begin{equation}
[T_{m},T_{n}] = f_{mn}{}^{p}T_{p} = -\epsilon_{mn}{}^{p} T_{p}\, ,  
\end{equation}

\noindent
so the SO$(3)$ generators in the fundamental/adjoint representation are the
matrices

\begin{equation}
T_{m}{}^{n}{}_{p} = \epsilon_{m}{}^{n}{}_{p} = -\epsilon_{mpn}\, .  
\end{equation}

We also need to know that the coset space SL$(3,\mathbb{R})/$SO$(3)$ is a
symmetric space and the structure constants with mixed indices $f_{ma}{}^{b}$
provide a representation of SO$(3)$ acting on the SL$(3,\mathbb{R})/$SO$(3)$
indices $a,b,\cdots$:

\begin{equation}
T_{m}{}^{a}{}_{b} = f_{mb}{}^{a}\, .  
\end{equation}

As for the generators of SL$(2,\mathbb{R})\sim$Sp$(2,\mathbb{R})$ in the
fundamental representation $T_{\alpha}{}^{i}{}_{j}$ we just need to know the
property

\begin{equation}
T_{\alpha}{}^{k}{}_{[j}\epsilon_{i]k} \equiv T_{\alpha\, [ij]}=0\, ,  
\end{equation}

Let us consider now the field content. In Ref.~\cite{Andino:2016bwk} the scalars were
collectively denoted by $\phi^{x}$.  We are going to keep using that notation
for the time being. The vector fields carried indices $I,J,\ldots$ and they
must be replaced by composite indices $im, jn$ etc.~where $i,j,\ldots=1,2$ and
$m,n,\ldots=1,2,3$ are indices in the fundamental representations of
SL$(2,\mathbb{R})$ and SL$(3,\mathbb{R})$, respectively. The notation for the
2-forms is the same. In Ref.~\cite{Andino:2016bwk} the electric 3-forms carry an
index $a$ which is the upper component of a symplectic index denoted by
$i,j,\ldots$. In the case at hands, $a$ takes only one value: $1$ ($C^{1}$)
which will be sometimes omitted ($C$). The lower index $1$ is equivalent to an
upper index $2$: $C_{1}= \epsilon_{12}C^{2} = C^{2}$ and, therefore $(C^{i})=
\left(
  \begin{smallmatrix}
    C^{1} \\ C_{1} \\
  \end{smallmatrix}
\right) =\left(
  \begin{smallmatrix}
    C^{1} \\ C^{2} \\
  \end{smallmatrix}
\right)$. On the other hand, $C_{i}\equiv \epsilon_{ij}C^{j}$.

Finally, in order to find the values of the $d$-tensors for this theory it is
enough to compare the field strengths of this theory with those of the generic
ungauged theory constructed in Ref.~\cite{Andino:2016bwk}. Comparing
Eqs.~(\ref{eq:Fimungauged}),(\ref{eq:Hmungauged}) and (\ref{eq:Giungauged})
with

\begin{eqnarray}
F^{I} 
& = &
d A^{I}\, .
\\
& & \nonumber \\
H_{m} 
& = &
dB_{m} - d_{mIJ}F^{I}A^{J}\, ,
\\
& & \nonumber \\
G^{i} 
& = &
dC^{i} +d^{i}{}_{I}{}^{m}F^{I}B_{m} 
-\tfrac{1}{3} d^{i}{}_{I}{}^{m}d_{mJK} A^{I}F^{J}A^{K}\, .
\end{eqnarray}

\noindent
we conclude that the $d$-tensors can be constructed entirely in terms of the
U-duality invariant tensors
$\delta^{i}{}_{j},\epsilon_{ij},\delta^{m}{}_{n},\epsilon_{mnp}$:

\begin{equation}
\begin{array}{rcl}
d_{mIJ} 
& \rightarrow & 
d_{m\, in\, jp} = -\tfrac{1}{2}\epsilon_{mnp}\epsilon_{ij}\, ,\\    
& & \\
d^{i}{}_{I}{}^{m}
& \rightarrow &
d^{i}{}_{jn}{}^{m} = \delta^{i}{}_{j}\delta^{m}{}_{n}\, .
\end{array}
\end{equation}

The tensor $d^{mnp}$ is related to these by

\begin{equation}
\label{eq:relationdddd}
d^{i}{}_{(I|}{}^{m}d_{i|J)}{}^{n}
=
-2d^{mnp}d_{pIJ}\, ,
\,\,\,\,\,
\Rightarrow
\,\,\,\,\,
d^{mnp} = +\tfrac{1}{2}\epsilon^{mnp}\, .
\end{equation}

We can immediately use the results of Ref.~\cite{Andino:2016bwk} to determine the
form of the 5-form field strengths $\tilde{H}^{m}$ (dual to the $H_{m}$) and
the 6-forms $\tilde{F}_{im}$ (dual to the 2-forms $F^{im}$)\footnote{The
  explicit expressions of the field strengths for a generic 8-dimensional
  theory is only given up to the 6-forms in Ref.~\cite{Andino:2016bwk}.}. We can
also derive the the Bianchi identities satisfied by all of them and also by
the 7-form field strengths $K_{A}$ dual to the Noether current 1-forms of the
scalar $\sigma$-model $j^{(\sigma)}_{A}$ where $A=m,a,\alpha$ runs in the
adjoint of the U-duality group. The later are given by

\begin{equation}
\label{eq:noetherdef}
j^{(\sigma)}_{A} \equiv \mathcal{G}_{xy} k_{A}{}^{x}d\phi^{y}\, ,  
\end{equation}

\noindent
where $\mathcal{G}_{xy}(\phi)$ is the $\sigma$-model metric and
$k_{A}{}^{x}(\phi)$ is the Killing vector of that metric associated to the
generator of the U-duality group $T_{A}$

\begin{equation}
[T_{A},T_{B}] = f_{AB}{}^{C}T_{C}\, ,
\hspace{1cm}
[k_{A},k_{B}] = -f_{AB}{}^{C}k_{C}\, .
\end{equation}

We are, however, interested in the gauged theory. The most general gaugings
can be found using the embedding-tensor formalism. In this theory, the
embedding tensor has the form $\vartheta_{im}{}^{A}$. We know there are at
least two possible SO$(3)\subset\mathrm{SL}(3,\mathbb{R})$ gaugings of this
theory:

\begin{enumerate}
\item Salam and Sezgin's \cite{Salam:1984ft}, in which the 3 vector fields
  $A^{1m}$ coming from the metric of 11-dimensional supergravity (that is,
  the 3 Kaluza-Klein (KK) vector fields) are used as gauge fields.
\item The AAMO \cite{AlonsoAlberca:2000gh} gauging in which the 3 gauge fields
  are the $A^{2m}$ coming from the 3-form of 11-dimensional supergravity.
\end{enumerate}

These two sets of gauge fields are related by the discreet electric-magnetic
SL$(2,\mathbb{R})$ duality transformation $\tau\rightarrow -1/\tau$ before
gauging. Correspondingly, the SS gauging corresponds to choosing an embedding
tensor whose only non-vanishing components are $\vartheta_{im}{}^{n}=
g\delta_{i}{}^{1}\delta_{m}{}^{n}$ where $g$ is the coupling constant, and the
AAMO gauging corresponds to the choice $\vartheta_{im}{}^{n}=
g\delta_{i}{}^{2}\delta_{m}{}^{n}$.

From the 8-dimensional supergravity point of view, one could use any other
SL$(2,\mathbb{R})$ transformed of the $A^{1m}$ triplet as gauge fields. This
suggests that a continuous family of equivalent SO$(3)$ gaugings should
exist. The corresponding embedding tensor has the form

\begin{equation}
\label{eq:embeddingtensor}
\vartheta_{im}{}^{n}= v_{i}\delta_{m}{}^{n}\, ,
\end{equation}

\noindent
where $v_{i}$ is a 2-component vector transforming in the fundamental of the
electric-magnetic SL$(2,\mathbb{R})$ duality group and can describe a
one-parameter family of equivalent SO$(3)$ gaugings of the theory\footnote{One
  of the two degrees of freedom of $v_{i}$ corresponds to the gauge coupling
  constant.}. The SO$(3)$ gauge fields are combinations of the two triplets of
vector fields

\begin{equation}
\vartheta_{in}{}^{m}A^{in}= v_{i}A^{im}\, ,  
\end{equation}

\noindent
and include, as limiting cases, the Salam-Sezgin and the AAMO theories.

Our candidate to embedding tensor Eq.~(\ref{eq:embeddingtensor}) must solve
the quadratic constraint, which implies its own gauge invariance

\begin{equation}
\vartheta_{im}{}^{B} Y_{B\, jn}{}^{A}= 0\, ,
\end{equation}

\noindent
where the $Y$ tensor is

\begin{equation}
  \begin{array}{rcl}
Y_{B\, jn}{}^{A} 
\equiv 
\delta_{B} \vartheta_{jn}{}^{A} 
& = &
-T_{B}{}^{k}{}_{j}\vartheta_{kn}{}^{A}
-T_{B}{}^{p}{}_{n}\vartheta_{jp}{}^{A}
+T_{B}{}^{A}{}_{C}\vartheta_{jn}{}^{C}
\\  
& & \\
& = &
-T_{B}{}^{k}{}_{j}\vartheta_{kn}{}^{A}
-T_{B}{}^{p}{}_{n}\vartheta_{jp}{}^{A}
+f_{BC}{}^{A}\vartheta_{jn}{}^{C}\, .
\end{array}
\end{equation}

For the above embedding tensor Eq.~(\ref{eq:embeddingtensor}), the only
non-vanishing components of the $Y_{A\, im}{}^{B}$ tensor are

\begin{equation}
\begin{array}{rclrcl}
Y_{a\, im}{}^{n} & = & -v_{i}T_{a}{}^{n}{}_{m}\, ,\hspace{1cm} 
&
Y_{a\, im}{}^{b} & = & -v_{i}f_{ma}{}^{b}\, ,
\\
& & & & & \\
Y_{\alpha\, im}{}^{n} & = & -T_{\alpha}{}^{j}{}_{i} v_{j} \delta_{m}{}^{n}\, ,
& & &  \\
\end{array}
\end{equation}

\noindent
and, therefore, the quadratic constraint is automatically satisfied and the
embedding tensor is, in principle, admissible.

There are other parameters associated to deformations of the theory that must
be considered together with the embedding tensor because they can be related.
The $d$-tensors being defined already in the undeformed theory, the rest of
the deformations of the theory are dictated by the St\"uckelberg mass
parameters $Z^{imn}$ and $Z_{im}$.

$Z^{imn}$ is related to the embedding tensor through the defining
relation\footnote{In this equation the parenthesis indicates the simultaneous
  symmetrization of the pairs of indices $im$ and $kp$.}

\begin{equation}
\vartheta_{(im|}{}^{A}T_{A}{}^{jn}{}_{|kp)} = Z^{jn q}d_{q\, im\, jp}\, .   
\end{equation}

\noindent
through the orthogonality relation

\begin{equation}
\vartheta_{im}{}^{A}Z^{imn}=0\, ,  
\end{equation}

\noindent
and through the requirement of gauge invariance

\begin{equation}
\vartheta_{im}{}^{A}Y_{A}{}^{jnp}=0\, ,
\,\,\,\,\,
\mbox{where}
\,\,\,\,\,
Y_{A}{}^{jnp} 
\equiv 
\delta_{A}Z^{inp}\, .
\end{equation}

\noindent
It is not difficult to see that the only solution to these three constraints
is

\begin{equation}
Z^{imn}= v^{i}\delta^{mn}\, ,
\hspace{1cm}
(v^{i}=\epsilon^{ji}v_{j})\, ,  
\end{equation}

\noindent
and, therefore, the only non-vanishing components of the tensor $Y_{a}{}^{imn}$
are

\begin{equation}
Y_{a}{}^{imn}  =  v^{i}T_{a}{}^{(m}{}_{q}\delta^{n)q}\, , 
\hspace{1cm}
Y_{\alpha}{}^{imn} =  T_{\alpha}{}^{i}{}_{j} v^{j} \delta^{mn}\, .
\end{equation}

$Z_{im}$ must be orthogonal to $Z^{inm}$

\begin{equation}
Z_{im}Z^{jnm}=0\, ,  
\end{equation}

\noindent
which can only be satisfied by $Z_{im}=0$. This solution is gauge-invariant
and the corresponding $Y$ tensor vanishes identically:

\begin{equation}
Y_{A\, im}=0\, ,
\hspace{1cm}
A=m,a,\alpha\, .
\end{equation}

There are five constraints more relating the three deformation tensors
$\vartheta_{im}{}^{A},Z^{imn}$ and $Z_{im}$ among themselves and to the
$d$-tensors \cite{Andino:2016bwk}:

\begin{equation}
\begin{array}{rcl}
\vartheta_{im}{}^{A}T_{A}{}^{p}{}_{n} +2d_{n\, im\, jq}Z^{jqp} +Z_{jn}d^{j}{}_{im}{}^{p} 
& = & 
0\, ,
\\
& & \\
\vartheta_{im}{}^{A}T_{A\, jk} +2Z_{(j|n}d_{|k)im}{}^{n}
& = &
0\, ,
\\
& & \\
d^{i}{}_{jp}{}^{[m|}Z^{jp|n]}+Z^{i}{}_{p}d^{pmn}
& = &
0\, ,
\\
& & \\
\tfrac{1}{2}d^{k}{}_{(ip|}{}^{m}d_{k|jq)}{}^{n}
+d^{mnp}d_{p\, ip\, jq} +3 d^{[m|}{}_{ip\, jq\, lr}Z^{lr|n]}
& = &
0\, ,
\\
& & \\
Z_{im}d^{m}{}_{jn\, kp\, lq} -d_{i(jn|}{}^{m}d_{m|kp\, lq)}
& = &
0\, ,
\\
\end{array}
\end{equation}

\noindent
where $d^{m}{}_{jn\, kp\, lq}$ is another $d$-tensor fully symmetric in the
three lower (pairs of) indices. They are satisfied identically when this
tensor vanishes.

The conclusion of this section is that we have found a set of deformation
parameters 

\begin{equation}
\vartheta_{im}{}^{n}= v_{i}\delta_{m}{}^{n}\, ,\hspace{.3cm}
Z^{imn}= v^{i}\delta^{mn}\, ,\hspace{.3cm}
Z_{im} =0\,   
\end{equation}

\noindent
that describe a one-parameter family of SO$(3)$ gaugings of maximal
8-dimensional supergravity with the properties we were looking for.

In what follows we are going to construct explicitly this family of theories
using the general results of Ref.~\cite{Andino:2016bwk}.

\section{Construction of the 1-parameter family of equivalent 
SO$(3)$-gauged  $\mathcal{N}=2$, $d=8$ supergravities}
\label{sec-construction}

The first step is the construction of the tensor hierarchy. Since this has
been done in Ref.~\cite{Andino:2016bwk} for most of the fields in a generic
8-dimensional theory, we just have to replace the values of the $d$-tensors
and the deformation tensors to get most of the field strengths and all the
Bianchi identities and the identities relating all the Bianchi
identities. They can be found in Appendices~\ref{sec-fieldstrengths},
\ref{sec-bianchiidentities} and~\ref{app-identitiesofBianchiidentities},
respectively. Nevertheless, we would like to remark the definitions of the
first and second covariant derivatives of the scalars

\begin{eqnarray}
\mathcal{D}\phi^{x} 
& \equiv &
d\phi^{x}-A^{im}v_{i}k_{m}{}^{x}\, ,
\\
& & \nonumber \\
\mathcal{D}\star \mathcal{D}\phi^{x} 
& \equiv &
d\star \mathcal{D}\phi^{x} 
+\Gamma_{yz}{}^{x}\mathcal{D}\phi^{y}\wedge \star \mathcal{D}\phi^{z}
-A^{im}v_{i}\partial_{y}k_{m}{}^{x} \wedge \star \mathcal{D}\phi^{y}\, ,
\end{eqnarray}

\noindent
and the fact that the Noether current 1-forms defined in
Eq.~(\ref{eq:noetherdef})\footnote{In absence of interactions between the
  scalars and other fields these Noether currents are conserved $d\star
  j_{A}^{(\sigma)}=0$. After gauging, in general they are no longer
  conserved. Their covariant generalizations are covariantly conserved
  $\mathcal{D}\star j_{A}^{(\sigma)}=0$, though. See
  Eq.~(\ref{eq:covariantnoetheredientity}) and its ungauged limit.} also need
to be covariantized

\begin{eqnarray}
\label{eq:noethercovariantdef}
j_{A}^{(\sigma)} 
& \equiv & 
k_{A}{}^{x}\mathcal{G}_{xy}\mathcal{D}\phi^{y}\, ,
\\
& & \nonumber \\
\mathcal{D}\star j_{A}^{(\sigma)}
& \equiv &
d\star j_{A} +f_{AB}{}^{C}A^{B}\wedge \star j_{C}\, .
\end{eqnarray}

As explained in Ref.~\cite{Andino:2016bwk}, the Bianchi identities of the magnetic
fields are related to the equations of motion of the electric ones upon the
use of the duality relations between electric and magnetic field strength
given in Appendix~\ref{sec-duality} and assuming that the Bianchi identities
of the electric field strengths are satisfied. The precise relation can be
found by studying the Noether identities associated to the gauge invariance of
the action of the theory (whose existence we assume) and, adapted to this
theory, is

\begin{eqnarray}
k_{A}{}^{x}\frac{\delta S}{\delta \phi^{x}} 
& = &
\mathcal{B}(K_{A})\, ,
\hspace{1cm}
A=m,a,\alpha\, ,
\\
& & \nonumber \\  
\frac{\delta S}{\delta A^{im}}
& = &
\mathcal{B}(\tilde{F}_{im})
+\left(
\delta^{1}{}_{i}B_{m} 
-\tfrac{1}{6}\epsilon_{mnp}A^{1n}A^{jp} 
\right)\mathcal{B}(G^{2})
-\tfrac{1}{2}\epsilon_{mnp}\epsilon_{ij}A^{jn}\mathcal{B}(\tilde{H}^{p})\, ,
\\
& & \nonumber \\  
\frac{\delta S}{\delta B_{m}}
& = & 
\mathcal{B}(\tilde{H}^{m})\, ,
\\
& & \nonumber \\  
\frac{\delta S}{\delta C^{1}}
& = &
\mathcal{B}(G^{2})\, .
\end{eqnarray}

From these relations we find

\begin{eqnarray}
k_{m}{}^{x}\frac{\delta S}{\delta \phi^{x}} 
& = &
-\mathcal{D}\star j^{(\sigma)}_{m}
-\epsilon_{m}{}^{n}{}_{p}
\left[
\mathcal{M}_{nq}\mathcal{W}_{ij}F^{ip}\wedge \star F^{jq} 
-\mathcal{M}^{pq}H_{n}\wedge \star H_{q}
\right]\, ,
\\
& & \nonumber \\  
k_{a}{}^{x}\frac{\delta S}{\delta \phi^{x}} 
& = &
-\mathcal{D}\star j^{(\sigma)}_{a} 
-T_{a}{}^{n}{}_{p}
\left[
\mathcal{M}_{nq}\mathcal{W}_{ij}F^{ip}\wedge \star F^{jq} 
-\mathcal{M}^{pq}H_{n}\wedge \star H_{q}
\right.
\nonumber \\
& & \nonumber \\  
& & 
\left.
-v_{i}\star \frac{\partial V}{\partial \vartheta_{ip}{}^{n}}
+v^{i}\star \frac{\partial V}{\partial Z_{ipn}}
\right]\, ,
\\
& & \nonumber \\  
k_{\alpha}{}^{x}\frac{\delta S}{\delta \phi^{x}} 
& = &
-d\star j^{(\sigma)}_{\alpha}
-T_{\alpha}{}^{i}{}_{j}
\left[
\mathcal{M}_{mn}\mathcal{W}_{ik}F^{jm}\wedge \star F^{kn} 
+\tfrac{1}{2}\mathcal{W}_{ik} G^{j} \wedge \star G^{k} 
\right.
\nonumber \\
& & \nonumber \\  
& & 
\left.
-v_{i}\delta^{m}{}_{n} \star \frac{\partial V}{\partial \vartheta_{jn}{}^{m}}
+v^{j}\delta^{mn} \star \frac{\partial V}{\partial Z_{imn}}
\right]\, ,
\\
& & \nonumber \\  
\frac{\delta S}{\delta A^{im}}
& = &
-\mathcal{D}(\mathcal{W}_{ij}\mathcal{M}_{mn}\star F^{jn})
-\epsilon_{mnp}\epsilon_{ij} F^{jn}\mathcal{M}^{pq}\star H_{q}
-(\delta_{i}{}^{1}\tilde{G}-\delta_{i}{}^{2}G)H_{m}
\nonumber \\
& & \nonumber \\  
& & 
-v_{i}K_{m}
+\left(
\delta^{1}{}_{i}B_{m} 
-\tfrac{1}{6}\epsilon_{mnp}A^{1n}A^{jp} 
\right)\frac{\delta S}{\delta C}
-\tfrac{1}{2}\epsilon_{mnp}\epsilon_{ij}A^{jn}\frac{\delta S}{\delta B_{p}}\, ,
\\
& & \nonumber \\  
\label{eq:eomBm}
\frac{\delta S}{\delta B_{m}}
& = & 
-\mathcal{D}(\mathcal{M}^{mn}\star H_{n}) +F^{1m}\tilde{G}- F^{2m}G 
+\tfrac{1}{2}\epsilon^{mnp}H_{n}H_{p} 
+v^{i}\mathcal{W}_{ij}\mathcal{M}_{mn}\star F^{jn}\, ,
\\
& & \nonumber \\  
\label{eq:eomC}
\frac{\delta S}{\delta C}
& = &
-d\tilde{G} +F^{2m}H_{m}\, .  
\end{eqnarray}

The scalar equations of motion can be recovered from the above three relations
by using 

\begin{enumerate}
\item The relation that expresses the gauge-invariance of the scalar potential

\begin{equation}
\label{eq:Vgaugeinvariance}
k_{A}{}^{x}\frac{\partial V}{\partial \phi^{x}} 
= 
Y_{A}{}^{\sharp}\frac{\partial V}{\partial c^{\sharp}}\, ,  
\end{equation}

\noindent
where the index $\sharp$ labels the deformations $c^{\sharp}$, which, in this
case, are just $\vartheta_{im}{}^{A},Z^{imn}$ and $Z_{im}$. Using the values
of the $Y$-tensors computed before and

\begin{equation}
\frac{\partial V}{\partial Z_{im}}=0\, ,  
\end{equation}

\noindent
we get the relations

\begin{eqnarray}
k_{m}{}^{x}\frac{\partial V}{\partial \phi^{x}} 
& = & 
0\, ,
\\
& & \nonumber \\
k_{a}{}^{x}\frac{\partial V}{\partial \phi^{x}} 
& = & 
-v_{i} T_{a}{}^{p}{}_{n}\frac{\partial V}{\partial \vartheta_{in}{}^{p}}
+v^{i} T_{a}{}^{n}{}_{p} \frac{\partial V}{\partial Z^{inp}}\, ,
\\
& & \nonumber \\
k_{\alpha}{}^{x}\frac{\partial V}{\partial \phi^{x}} 
& = & 
-T_{\alpha}{}^{j}{}_{i}v_{j}\delta^{p}{}_{n} 
\frac{\partial V}{\partial \vartheta_{in}{}^{p}}
+T_{\alpha}{}^{i}{}_{j}v^{j}\delta^{np}
\frac{\partial V}{\partial Z_{inp}}\, .
\end{eqnarray}

\item The invariance of the theory under the U-duality group implies that the
  kinetic matrices $\mathcal{M}_{mn}(\phi)$ and $\mathcal{W}_{ij}$ satisfy the
  following relations:

\begin{equation}
  \begin{array}{rcl}
k_{m}{}^{x}\partial_{x}\mathcal{M}_{np} 
& = & 
-2\epsilon_{m}{}^{q}{}_{(n}\mathcal{M}_{p)q}\, ,
\\
& & \\
k_{a}{}^{x}\partial_{x}\mathcal{M}_{np} 
& = & 
-2T_{a}{}^{q}{}_{(n}\mathcal{M}_{p)q}\, ,
\\
& & \\
k_{\alpha}{}^{x}\partial_{x}\mathcal{W}_{ij} 
& = & 
-2T_{\alpha}{}^{K}{}_{(i}\mathcal{W}_{j)k}\, .
\end{array}
\end{equation}

The axidilaton field $\tau$ transforms non-linearly under SL$(2,\mathbb{R})$
(fractional-linear transformations). Taking into account the (unconventional,
by an overall sign) definition of the dual 4-form $\tilde{G}$ that constitutes
the lower entry of the symplectic vector $G^{i}$\footnote{It is this definition
  that brings us to the unconventional SL$(2,\mathbb{R})$ matrix
  $\mathcal{W}$}, the infinitesimal SL$(2,\mathbb{R})$ transformations of
$\tau$ take the form

\begin{equation}
k_{\alpha}{}^{x}\partial_{x} \tau
=
-T_{\alpha\, 11} +(T_{\alpha\, 1}{}^{1} -T_{\alpha}{}^{1}{}_{1}) \tau
+T_{\alpha}{}^{11}\tau^{2}\, .  
\end{equation}

\item Finally, using the Killing equation it is not difficult to prove the
  following identity for the Killing vectors $k_{A}{}^{x}$ of a metric
  $\mathcal{G}_{xy}(\phi)$ and the associated covariantized Noether 1-form
defined in Eq.~(\ref{eq:noethercovariantdef})

\begin{equation}
\label{eq:covariantnoetheredientity}
k_{A\, x}
\mathcal{D}\star \mathcal{D}\phi^{y}
=
\mathcal{D}\star j^{(\sigma)}_{A}\, .
\end{equation}

\end{enumerate}

Then, the scalar equations of motion are 

\begin{equation}
  \begin{array}{rcl}
{\displaystyle
\frac{\delta S}{\delta \phi^{y}} 
}
& = &
-\mathcal{G}_{xy}\mathcal{D}\star \mathcal{D}\phi^{y} 
+\tfrac{1}{2}\partial_{x}
\left\{
\mathcal{W}_{ij}\mathcal{M}_{mn}F^{im}\wedge \star F^{jn}
+\mathcal{M}^{mn} H_{m}\wedge \star H_{n}
\right.
\\
& & \\
& & 
\left.
+e^{-\varphi} G \wedge \star G
- a G\wedge  G
-V(\phi)
\right\}\, .  
\end{array}
\end{equation}

These equations can be split into those corresponding to the scalars in the
coset spaces SL$(3,\mathbb{R})/\mathrm{SO}(3)$
and SL$(2,\mathbb{R})/\mathrm{SO}(2)$ in the obvious way.

We will discuss the form of the potential later.

The scalar equations of motion give us all the kinetic terms in the action:

\begin{equation}
\label{eq:S0}
\begin{array}{rcl}
S^{(0)} 
& = & 
{\displaystyle\int} 
\left\{ 
-\star R
+\frac{1}{4}{\rm Tr}
\left(
\mathcal{D} \mathcal{M}\mathcal{M}^{-1}
\wedge 
\star \mathcal{D} \mathcal{M}\mathcal{M}^{-1} \right)
+\frac{1}{4}{\rm Tr}
\left(
d \mathcal{W}\mathcal{W}^{-1}
\wedge 
\star d \mathcal{W}\mathcal{W}^{-1} \right)
\right.
\\
& & \\
& & 
\left.
+\frac{1}{2}\mathcal{W}_{ij}\mathcal{M}_{mn}F^{im}\wedge \star F^{jn}
+\frac{1}{2} \mathcal{M}^{mn} H_{m}\wedge \star H_{n}
+\frac{1}{2} e^{-\varphi} G \wedge \star G
-\frac{1}{2} a G\wedge G
-V
\right\} \, .
\end{array}
\end{equation}

(We have added the Hilbert-Einstein term, which, evidently, should be there).
Now we have to add the Chern-Simons terms necessary to obtain the other
equations of motion, starting by those of the higher-rank potentials
($C$). However, all the Chern-Simons terms of the ungauged theory must be
present (since we must recover it in the $v^{i}=0$ limit) and it makes sense
to add to the above action the covariantization of those terms, namely

\begin{equation}
\label{eq:S1}
\begin{array}{rcl}
S^{(1)} 
& = & 
{\displaystyle\int} 
\left\{ 
-dC^{1}\Delta G^{2} -\tfrac{1}{2}\Delta G^{1}\Delta G^{2}
-\tfrac{1}{12}\epsilon^{mnp}B_{m}\mathcal{D}B_{n}\mathcal{D}B_{p}
+\tfrac{1}{4}\epsilon^{mnp}B_{m}H_{n}H_{p}
\right. 
\\
& & \\
& & 
\left.
-\tfrac{1}{24}\epsilon_{ij} A^{im} A^{jn} \Delta H_{m}\mathcal{D}B_{n}
\right\} \, ,
\\
\end{array}
\end{equation}

\noindent
where now the field strengths and derivatives are the covariant ones and 

\begin{equation}
\Delta H_{m}= H_{m}-\mathcal{D}B_{m}\, ,
\hspace{1cm}
\Delta G^{i} = G^{i}- dC^{i}\, .
\end{equation}

$C$ only occurs in one place in this Chern-Simons term and, therefore, using
$d\Delta G^{2} = dG^{2}$ and the Bianchi identity $\mathcal{B}(G^{2})$ in
  Eq.~(\ref{eq:BianchiGi}) we get 

\begin{equation}
\frac{\delta S^{(0)}+S^{(1)}}{\delta C^{1}} 
= -d\tilde{G} +d\Delta G^{2} = -d\tilde{G} +F^{2m}H_{m}\, ,  
\end{equation}

\noindent
in agreement with Eq.~(\ref{eq:eomC}).

For the 2-forms we find 

\begin{equation}
\begin{array}{rcl}
{\displaystyle
\frac{\delta S^{(0)}+S^{(1)}}{\delta B_{m}} 
}
& = &
{\displaystyle
\frac{\delta S}{\delta B_{m}}
}
+\tfrac{1}{12}v_{i}F^{im}B_{n}B_{n}
+\tfrac{1}{6}v_{i}F^{in}B_{n}B_{m}
+\tfrac{1}{2}v_{i}\Box G^{i}B_{m}
\\
& & \\
& & 
+\tfrac{1}{2}\epsilon_{ij}\Box G^{i} \Box F^{jm} 
-\tfrac{1}{4}\epsilon^{mnp}\Delta H_{n} \Delta H_{p}
+\tfrac{1}{24}\mathcal{D}\left(\epsilon_{ij}A^{im}A^{in}\Delta H_{n}\right)
\, ,  
\end{array}
\end{equation}

\noindent
where $\delta S/\delta B_{m}$ is the expected equation of motion, given in
Eq.~(\ref{eq:eomBm}), and where the boxes acting on field strengths denote the
terms on that field strength that only depend on the 1-form fields. Thus, the
terms in the second line only depend on the 1-form fields and it is very easy
to add a term to the action, linear in $B_{m}$ to cancel them. However, we
must make sure, first, that those terms always depend on $v^{i}$, so they
disappear in the ungauged limit. Indeed, expanding them we find that all the
$v$-independent terms in them cancel. As for the unwanted terms in the first
line (all of them $v$-dependent), they can be easily integrated. We conclude
that we must add to the action a new correction:

\begin{equation}
  \begin{array}{rcl}
S^{(2)} 
& = &
{\displaystyle\int}
\left\{
-\tfrac{1}{12}v_{i}(F^{im}- v^{i}B_{m})B_{m}B_{n}B_{n}
+\tfrac{1}{4} \epsilon^{mnp}B_{m}\Delta H_{n} \Delta H_{p}
-\tfrac{1}{2}\epsilon_{ij}\Box G^{i} \Box F^{jm}B_{m}
\right.
\\
& & \\
& & 
\left.
+\tfrac{1}{24}\epsilon_{ij}A^{im}A^{in}\mathcal{D}B_{m}\Delta H_{n}
\right\}\, .   
\end{array}
\end{equation}

\noindent
Varying $S^{(0)}+S^{(1)}+S^{(2)}$ with respect to $C$ and $B_{m}$ gives the
expected equations of motion. 

The terms that remain to be added only contain 1-forms and their derivatives
and only contribute to the equations of motion of the 1-forms. They are of the
form $(dA)^{2}A^{4}$ and $(dA)A^{6}$ and their form is exceedingly complicated
and we have not determined them.

\subsection{The scalar potential}

Finally, we have to find the scalar potential. The scalar potential must
satisfy Eq.~(\ref{eq:Vgaugeinvariance}), but this equation does not fully
determine it. In supergravity theories, the scalar potential is determined by
supersymmetry, and is quadratic in the \textit{fermion shifts}.\footnote{The
  exception is $\mathcal{N}=1,d=4$ supergravity, which, even in the ungauged
  case, admits a scalar potential entirely built from the superpotential,
  which is largely arbitrary.}

There seem to be no general rules available in the literature to construct the
fermion shifts of any gauged supergravity, although, based on the example of
gauge $\mathcal{N}=3,d=4$ supergravity \cite{Castellani:1985ka}, it was
suggested in Ref.~\cite{Castellani:1985wk} that they can be written in terms
of the \textit{dressed structure constants} of the gauge group.

Looking into Ref.~\cite{Salam:1984ft}, we can see that the fermion shifts of
SO$(3)$-gauged $\mathcal{N}=2,d=8$ supergravity theory fit into this general
rule and are written in terms of 

\begin{equation}
f_{\mathbf{m}\mathbf{n}}{}^{\mathbf{p}} 
\equiv 
L_{\mathbf{m}}{}^{m}  L_{\mathbf{n}}{}^{n} L_{p}{}^{\mathbf{p}}f_{mn}{}^{p}\, ,
\end{equation}

\noindent
where $f_{mn}{}^{p}=\epsilon_{mnp}$, the matrix $L_{\mathbf{m}}{}^{n}$ is the
SL$(3,\mathbb{R})/$SO$(3)$ coset representative, and $L_{m}{}^{\mathbf{n}}$is
its inverse.\footnote{Here $m,n,p=1,2,3$ are, as in the rest of this paper,
  indices of the fundamental (vector) representation of SL$(3,\mathbb{R})$
  and $\mathbf{m},\mathbf{n},\mathbf{p}=1,2,3$ are indices in the fundamental
  representation of SO$(3)$).}

It is, however, well-known that in $\mathcal{N}=1,2,d=4$ supergravities the
fermion shifts are written in terms of the \textit{momentum maps}
$P_{A}{}^{\Sigma}$ associated to the symmetries being gauged: the index $A$
runs over its Lie algebra and the index $\Sigma$ runs over the Lie Algebra of
the R-symmetry group. Thus, in this theory, they would have be $P_{A}{}^{m}$
with $A=m,a,\alpha$.

As discussed in Ref.~\cite{Bandos:2016smv}, these two ways of writing fermion shifts
are, actually, equivalent because the dressed structure constants can be
rewritten in terms of the momentum maps. The momentum maps, though, can be
combined with the embedding tensor in a natural way
($\vartheta_{im}{}^{n}P_{n}{}^{p}$) and more general gaugings can be
considered. We will, therefore, use the momentum maps to write the fermion
shifts of the theory at hands.

A problem one finds in trying to write fermion shifts with the right structure
is that the structure of the fermion shifts and of the entire supersymmetry
transformations given in Ref.~\cite{Salam:1984ft} does not show the
transformation properties of the spinors under the R-symmetry group
SO$(2)\times$SO$(3)\sim$U$(1)\times$SU$(2)$, because the fermions obtained in
the dimensional reduction from 11 dimensions are not symplectic-Majorana. A
symplectic-Majorana (pair) $\epsilon^{I}$ $I=1,2$ transforms as a doublet
under SU$(2)$ and as a singlet under U$(1)$ in a natural way. Therefore, we
are going to use symplectic-Majorana spinors in our proposal: gravitini
$\psi_{\mu I}$, dilatini $\lambda_{m}^{I}$ and supersymmetry parameters
$\epsilon^{I}$ and we are going to define the fermion shifts $S_{IJ},
N_{\mathbf{m}}{}^{I}{}_{J}$

\begin{equation}
\begin{array}{rcl}
\delta_{\epsilon}\psi_{\mu I} 
& \sim & 
\cdots +S_{IJ}\epsilon^{J}\, , \\
& & \\
\delta_{\epsilon}\lambda_{\mathbf{m}}^{I} 
& \sim & 
\cdots +N_{\mathbf{m}}{}^{I}{}_{J}\epsilon^{J}\, . \\
\end{array}
\end{equation}

Now, in order to construct $S_{IJ}$ and $N_{m}{}^{I}{}_{J}$ it is necessary to
introduce an object with properties similar to those of the symplectic
sections of $\mathcal{N}=2,d=4$ theories and their generalizations to higher
$\mathcal{N}$ denoted by $\mathcal{V}^{M}{}_{IJ}$ where the index $M$ labels
the vectors available in the theory (electric and magnetic in 4 dimensions)
and the indices $I,J$ are R-symmetry indices in the representation carried by
the spinors (the fundamental of SU$()\mathcal{N}$). This generalization, must
have the same structure, i.e. $\mathcal{V}^{im}{}_{IJ}$ and our proposal for
this object is

\begin{equation}
  \mathcal{V}^{im}{}_{IJ} \equiv V^{i}L_{\mathbf{m}}{}^{m} 
\epsilon_{IK}\sigma^{\mathbf{m}\, K}{}_{J}\, , 
\,\,\,\,\,
\mbox{and}
\,\,\,\,\,
  \mathcal{V}^{im}{}_{\mathbf{m}} \equiv V^{i}L_{\mathbf{m}}{}^{m}\, , 
\end{equation}

\noindent
where we have introduced 

\begin{equation}
(V_{i}) \equiv e^{\varphi/2}(\tau\,\,\,\,\, 1)\, ,  
\end{equation}

\noindent
which transforms linearly under SL$(2,\mathbb{R})$ up to a U$(1)$ phase.

Using these ingredients, the fermion shifts can be written in the form 

\begin{eqnarray}
S_{IJ} 
& = & 
\mathcal{V}^{im}{}_{[I|K}
\vartheta_{im}{}^{n}P_{n}{}^{\mathbf{p}}(\sigma^{\mathbf{p}})^{K}{}_{|J]}\, ,
\\
& & \nonumber \\
N_{\mathbf{m}}{}^{I}{}_{J}   
& = &
\mathcal{V}^{in}{}_{\mathbf{r}}
\vartheta_{in}{}^{p}P_{p}{}^{\mathbf{s}}
\left(\delta^{\mathbf{r}}{}_{\mathbf{m}}\delta^{\mathbf{q}}{}_{\mathbf{s}}
-\tfrac{1}{2}\delta_{\mathbf{m}}{}^{\mathbf{q}}
\delta^{\mathbf{r}}{}_{\mathbf{s}}
 \right)
(\sigma^{\mathbf{q}})^{I}{}_{J}\, ,
\end{eqnarray}

\noindent
where the $(\sigma^{\mathbf{p}})$ are Pauli's sigma matrices. For the class of
gaugings that we are considering, with embedding tensor
$\vartheta_{im}{}^{n}=v_{i}\delta_{m}{}^{n}$

\begin{eqnarray}
S_{IJ} 
& = & 
V^{i}v_{i}L_{\mathbf{n}}{}^{m}P_{m}{}^{\mathbf{n}}\epsilon_{IJ}\, ,
\\
& & \nonumber \\
N_{\mathbf{m}}{}^{I}{}_{J}   
& = &
V^{i}v_{i} 
\left(L_{\mathbf{m}}{}^{n}P_{n}{}^{\mathbf{p}}
-\tfrac{1}{2}\delta_{\mathbf{m}}{}^{\mathbf{p}}
L_{\mathbf{q}}{}^{n}P_{n}{}^{\mathbf{q}}
 \right)
(\sigma^{\mathbf{p}})^{I}{}_{J}
\, .
\end{eqnarray}

\noindent
Now we observe that the dressed structure constants can, in this case, be
expressed in these two different ways:

\begin{equation}
  f_{\mathbf{m}\mathbf{n}}{}^{\mathbf{p}} 
=
\left\{
  \begin{array}{l}
  L_{\mathbf{m}}{}^{q}\Gamma_{\rm Adj}(L^{-1})_{q}{}^{A}
(T_{A})_{\mathbf{n}}{}^{\mathbf{p}}\, ,\\
\\
\epsilon_{\mathbf{m}\mathbf{n}\mathbf{q}} T^{\mathbf{q}\mathbf{p}}\, ,\\
  \end{array}
\right.
\end{equation}

\noindent
where we have defined 

\begin{equation}
T^{\mathbf{m}\mathbf{n}} \equiv L_{p}{}^{\mathbf{m}}L_{p}{}^{\mathbf{n}}\, .
\end{equation}

\noindent
Contracting both identities with $\epsilon^{\mathbf{n}\mathbf{p}\mathbf{r}}$
we find\footnote{This projects the first identity over the SO$(3)$ generators
  $A=m$ and we remind the reader our definition of momentum map
  $P_{B}{}^{\mathbf{m}}= \Gamma_{\rm Adj}(L^{-1})_{B}{}^{\mathbf{m}}$.}

\begin{equation}
2L_{\mathbf{m}}{}^{p}P_{p}{}^{\mathbf{n}} 
=
-T^{\mathbf{m}\mathbf{n}}+\delta^{\mathbf{m}\mathbf{n}}T\, ,
\,\,\,\,\,
\mbox{where}
\,\,\,\,\,  
T\equiv \delta_{\mathbf{m}\mathbf{n}}T^{\mathbf{m}\mathbf{n}}\, ,
\end{equation}

\noindent
which allows us to express the fermion shifts entirely in terms of
$T^{\mathbf{m}\mathbf{n}}$:

\begin{eqnarray}
S_{IJ} 
& = & 
\epsilon_{IJ}V^{i}v_{i}T\, ,
\\
& & \nonumber \\
N_{\mathbf{m}}{}^{I}{}_{J}   
& = &
V^{i}v_{i} 
\left(T^{\mathbf{m}\mathbf{p}}
-\tfrac{1}{2}\delta_{\mathbf{m}\mathbf{p}}T
 \right)
(\sigma^{\mathbf{p}})^{I}{}_{J}
\, ,
\end{eqnarray}

\noindent
as they appear in Ref.~\cite{Salam:1984ft}. 

The combination of the fermion shifts that gives the scalar potential is 

\begin{equation}
V
= 
-\tfrac{1}{4}S_{IJ}S^{*\, IJ} +\tfrac{1}{8}\delta^{\mathbf{m}\mathbf{n}}
N_{\mathbf{m}}{}^{I}{}_{J}N^{*}_{\mathbf{n}}{}_{I}{}^{J}
=
-\tfrac{1}{2}\mathcal{W}^{ij}v_{i}v_{j} 
\left[\mathrm{Tr}(\mathcal{M})^{2}
-2\mathrm{Tr}(\mathcal{M}^{2})\right]\, ,     
\end{equation}
 
\noindent
where $\mathcal{W}^{ij}$ is the SL$(2,\mathbb{R})/\mathrm{SO}(2)$
symmetric matrix defined in Eq.~(\ref{eq:Wijdef}) and where we have used, to
simplify the comparison with the results of
Refs.~\cite{AlonsoAlberca:2000gh,AlonsoAlberca:2003jq}

\begin{equation}
\mathcal{M}_{mn}\equiv L_{m}{}^{\mathbf{p}}L_{n}{}^{\mathbf{p}}\, ,
\,\,\,\,\,
\mbox{so that}
\,\,\,\,\,
T=\mathrm{Tr}(\mathcal{M})\, ,
\,\,\,\,\,
\mbox{and}
\,\,\,\,\,
T^{\mathbf{m}\mathbf{n}}T^{\mathbf{m}\mathbf{n}}
=\mathrm{Tr}(\mathcal{M}^{2})\, .
\end{equation}

The expression obtained is SO$(3)$ invariant and, formally (because $v^{i}$ is
rotated) SL$(2,\mathbb{R})$ invariant. For $v_{i}= g\delta_{i}{}^{1}$ one
recovers the scalar potential of the Salam-Sezgin theory\footnote{Beware of
  the different conventions for the dilaton field!} and for $v_{i}=
g\delta_{i}{}^{2}$ the scalar potential of the AAMO theory is recovered.

Since our main concern here was to find the scalar potential, in this section
we have only studied the fermion shifts in the supersymmetry transformations
of the fermions. It is worth, however, discussing the general form of all the
supersymmetry transformations of the theory. We know that the supersymmetry
transformations of the bosonic fields do not change under gauging, as a
rule. The structure of the supersymmetry transformations of the fermions is,
apart from the additional fermion shifts, the same as in the ungauged case
with the field strengths replaced by the new ones and the derivatives replaced
by gauge-covariant derivatives. Since the fermions only transform under the
R-symmetry group, their covariant derivatives are different from the covariant
derivatives of the bosonic fields, which transform in representations of the
whole duality group. The construction of these covariant derivatives offers no
particular problems and is discussed in detail in Ref.~\cite{Bandos:2016smv}.

\section{Conclusions}
\label{sec-conclusions}

By applying the general results obtained in Ref.~\cite{Andino:2016bwk} we have
constructed explicitly a 1-parameter family of SL$(2,\mathbb{R})$-related,
SO$(3)$-gauged maximal 8-dimensional supergravities that interpolates between
Salam and Sezgin's \cite{Salam:1984ft} and AAMO's \cite{AlonsoAlberca:2000gh},
realizing the possibilities noticed in
Refs.~\cite{deRoo:2011fa,Dibitetto:2012rk}: for each value of that parameter a
different combination of the two triplets of 1-forms (one coming from the
11-dimensional metric and the other coming from the 11-dimensional 3-form)
plays the r\^ole of gauge vectors. 

The existence of this family confirms the identification of AAMO with a honest
SO$(3)$-gauged maximal 8-dimensional supergravity in spite of its very
unconventional origin: the so-called \textit{massive 11-dimensional
  supergravity} proposed in Ref.~\cite{Meessen:1998qm}. Furthermore, it proves
its relation with the Salam-Sezgin theory by an SL$(2,\mathbb{R})$
transformation, something that would have been very difficult to do directly.
Thus, we have achieved the two goals stated in the introduction. At the same
time, our result poses further questions: what is the 11-dimensional origin of
all the theories in this family if we insist in using the same
compactification Ansatz?

A key ingredient of the gauged supergravities we have constructed is the
scalar potential. This is not determined by the tensor hierarchy, which only
puts generic constraints on it. In a supergravity theory (different from
$\mathcal{N}=1,d=4$) the scalar potential is a quadratic form on the fermion
shifts. These have to be scalar-dependent expressions linear on the embedding
tensor, but their general form is not known.\footnote{They are
  known in $\mathcal{N}=2,d=4,5$ supergravity and in other $\mathcal{N}\geq 3,
  d=4$ supergravities as well. See Ref.~\cite{Bandos:2016smv} and references
  therein.} This is one of the main obstructions to find a general formulation
of all gauged supergravities in all dimensions. We have proposed a general
form of the fermion shifts for maximal 8-dimensional supergravity that
reproduces the fermion shifts found by Salam and Sezgin and gives the expected
(formally) duality-invariant form of the scalar. Interestingly enough, this
form is similar to that of the fermion shifts occurring in 4-dimensional
supergravities, where the scalar fields appear combined in an object
(symplectic section and generalizations) related to part of the coset
representative. We believe that this object should exist in any supergravity
theory (if it can be gauged at all) and its identification and study should be
the key for finding the general formulation of gauged supergravities we wish for.

\section*{Acknowledgments}

T.O.~would like to thank M.~Trigiante and Gianluca Inverso for useful
conversations during the workshop \textit{Superstring solutions, supersymmetry
  and geometry} held at the Centro de Ciencias de Benasque and
J.J.~Fern\'andez-Melgarejo for useful conversations on the determination of
the fermions shifts.  This work has been supported in part by the Spanish
Ministry of Science and Education grants FPA2012-35043-C02-01 and
FPA2015-66793-P, the Centro de Excelencia Severo Ochoa Program grant
SEV-2012-0249, and the Spanish Consolider-Ingenio 2010 program CPAN
CSD2007-00042.  The work of OLA was further supported by a scholarship of the
Ecuadorian Secretary of Science, Technology and Innovation.  TO wishes to
thank M.M.~Fern\'andez for her permanent support.

\appendix

\section{Summary of relations for the gauged theory}

\subsection{Field strengths}
\label{sec-fieldstrengths}

\begin{eqnarray}
\mathcal{D}\phi^{x}
& = & 
d\phi^{x} -A^{im}v_{i}k_{m}{}^{x}\, ,
\\
& & \nonumber \\
F^{im} 
& = &
dA^{im} +\tfrac{1}{2}\epsilon^{mnp}v_{j}A^{jn}A^{ip} +v^{i}B_{m}\, ,
\\
& & \nonumber \\
H_{m} 
& = &
\mathcal{D}B_{m} +\tfrac{1}{2}\epsilon_{mnp}\epsilon_{ij}dA^{in}A^{jp}
-\tfrac{1}{4}v_{i}A^{im}(\epsilon AA)\, ,
\\
& & \nonumber \\
G^{i} 
& = &
dC^{i} +F^{im}B_{m}
+\tfrac{1}{6}\epsilon_{mnp}\epsilon_{jk}dA^{jn}A^{im}A^{kp}
-v^{i}\left[\tfrac{1}{2}B_{m}B_{m}-\tfrac{1}{32}(\epsilon AA)^{2} \right]\, ,
\\
& & \nonumber \\
\tilde{H}^{m}
& = &
\mathcal{D}\tilde{B}^{m} +\epsilon_{ij}F^{im}C^{j}
+\tfrac{1}{2}\epsilon^{mnp}B_{n}\left(H_{p}+\Delta H_{p}\right)
\nonumber \\
& & \nonumber \\
& & 
+\tfrac{1}{24}\epsilon_{pqr} \epsilon_{ij}\epsilon_{kl}dA^{ip}A^{jq}A^{kr}A^{lm}
+\tfrac{1}{160}v_{i}A^{im}(\epsilon AA)^{2} +v^{i}\tilde{A}_{im}\, ,
\\
& & \nonumber \\
\tilde{F}_{im} 
& = &
\mathcal{D} \tilde{A}_{im} 
-\epsilon_{ij}\epsilon_{mnp}F^{jn}\tilde{B}_{p} 
-\epsilon_{ij}H_{m}C^{j}
-v_{i}B_{m}B_{n}B_{n} -\tfrac{1}{2}\epsilon_{ij}\Delta F^{jn}B_{m}B_{n}
\nonumber \\
& & \nonumber \\
& & 
+\tfrac{1}{24} \epsilon_{il}\epsilon_{jj^{\prime}}\epsilon_{kk^{\prime}}
\epsilon_{mnp}\epsilon_{qrs}dA^{jq} A^{j^{\prime}r}
A^{ks}A^{k^{\prime}n}A^{lp}
+\tfrac{1}{16}\epsilon_{ij}v_{k}\epsilon_{mnp}A^{jn}A^{kp}(\epsilon AA)^{2}
\nonumber \\
& & \nonumber \\
& & 
-v_{i}D_{m}\, ,
\\
& & \nonumber \\
K_{m}
& = &
\mathcal{D}D_{m} + \ldots\, ,
\\
& & \nonumber \\
K_{a}
& = &
\mathcal{D}D_{a} + \ldots\, ,
\\
& & \nonumber \\
K_{\alpha}
& = &
dD_{\alpha} + \ldots\, ,
\end{eqnarray}

\noindent
where the SO$(3)$-covariant derivatives that appear in these expressions are 

\begin{equation}
\label{eq:covariantderivativesBm}
\mathcal{D}B_{m} = dB_{m}+\epsilon_{mnp}v_{i}A^{in}B_{p}\, ,
\hspace{1cm}  
\mathcal{D}\tilde{B}^{m} = d\tilde{B}^{m}+\epsilon^{mnp}v_{i}A^{in}\tilde{B}^{p}\, .
\end{equation}

\noindent
and where we have used the shorthand notation

\begin{equation}
\Delta H_{m} = H_{m}-\mathcal{D}B_{m}\, ,
\hspace{1cm}
(\epsilon AA) = \epsilon_{ij}A^{im}A^{jn}\, .  
\end{equation}

\subsection{Bianchi identities}
\label{sec-bianchiidentities}

The Bianchi identities satisfied by the field strengths of the gauged theory
are $\mathcal{B}(\cdot)=0$ where

\begin{eqnarray}
\mathcal{B}(L_{n}{}^{im})
& = &
-\left[
\mathcal{D}L_{n}{}^{im} +F^{im}K_{n} +W^{im}{}_{n}{}^{\beta}M_{\beta}
\right]\, ,
\\
& & \nonumber \\
\mathcal{B}(L_{a}{}^{im})
& = &
-\left[
\mathcal{D}L_{a}{}^{im} +F^{im}K_{a} +W^{im}{}_{n}{}^{\beta}M_{\beta}
\right]\, ,
\\
& & \nonumber \\
\mathcal{B}(L_{imn})
& = &
-\left[
\mathcal{D}L_{imn} +2\tilde{F}_{i(m}H_{n)} +W_{imn}{}^{\beta}M_{\beta}
\right]\, ,
\\
& & \nonumber \\
\mathcal{B}(K_{m})
& = &
\mathcal{D}K_{m}
-T_{m}{}^{n}{}_{p}
\left[
F^{ip}\tilde{F}_{in} +\tilde{H}^{p}H_{n}
\right]\, ,
\\
& & \nonumber \\
\mathcal{B}(K_{a})
& = &
\mathcal{D}K_{a} 
-T_{a}{}^{n}{}_{p}
\left[
F^{ip}\tilde{F}_{in} +\tilde{H}^{p}H_{n}
-v_{i}L^{ip}{}_{n}+v^{i}\delta^{mp} L_{imn}
\right]\, ,
\\
& & \nonumber \\
\mathcal{B}(K_{\alpha})
& = &
dK_{\alpha}
-T_{\alpha}{}^{i}{}_{j}
\left[
F^{jm}\tilde{F}_{im} +\tfrac{1}{2}G^{j}G_{i} 
-v_{i}\delta^{m}{}_{n} L^{jn}{}_{m}+v^{j}\delta^{mn} L_{imn}
\right]\, ,
\\
& & \nonumber \\
\mathcal{B}(\tilde{F}_{im})
& = &
-\left[
\mathcal{D}\tilde{F}_{im}
+\epsilon_{mnp}\epsilon_{ij} F^{jn}\tilde{H}^{p}
+\epsilon_{ij}G^{j}H_{m}
+v_{i}K_{m}
\right]\, ,
\\
& & \nonumber \\
\mathcal{B}(\tilde{H}^{m})
& = &
-\left[
\mathcal{D} \tilde{H}^{m} 
-\epsilon_{ij}F^{im}G^{j} -\tfrac{1}{2}\epsilon^{mnp}H_{n}H_{p}
-v^{i}\tilde{F}_{im}
\right]\, ,
\\
& & \nonumber \\
\label{eq:BianchiGi}
\mathcal{B}(G^{i})
& = &
-\left[
dG^{i} 
-F^{im}H_{m}
\right]\, ,
\\
& & \nonumber \\
\mathcal{B}(H_{m})
& = &
-\left[
\mathcal{D}H_{m}
-\tfrac{1}{2}\epsilon_{mnp}\epsilon_{ij}F^{im}F^{jn}
\right]\, , 
\\
& & \nonumber \\
\mathcal{B}(F^{im})
& = &
-\left[
\mathcal{D}F^{im} -v^{i}H_{m}
\right]\, ,
\\
& & \nonumber \\
\mathcal{B}(\mathcal{D}\mathcal{M}_{mn})
& = &
-\left[
\mathcal{D}\mathcal{D}\mathcal{M}_{mn}
+2v_{i}F^{ip}\epsilon_{p(m}{}^{q}\mathcal{M}_{n)q}
\right]\, ,
\\
& & \nonumber \\
\mathcal{B}(d\mathcal{W}_{ij})
& = &
-dd\mathcal{W}_{ij}\, ,
\end{eqnarray}

\noindent
where the SO$(3)$-covariant derivatives with indices $m$ are identical to
those of $B_{m}$ and $\tilde{B}^{m}$ in Eq.~(\ref{eq:covariantderivativesBm}) 

\begin{equation}
\label{eq:covariantderivativeKa}
\mathcal{D}K_{a} = dK_{a} -v_{i}A^{im}f_{ma}{}^{b}K_{b}\, .  
\end{equation}

\subsection{Identities of Bianchi identities}
\label{app-identitiesofBianchiidentities}

\begin{eqnarray}
\mathcal{D}\mathcal{B}(F^{im}) -v^{i}\mathcal{B}(H_{m}) 
& = & 0\, ,
\\
& & \nonumber \\
\mathcal{D}\mathcal{B}(H_{m})+\epsilon_{ij}\epsilon_{mnp}F^{in}\mathcal{B}(F^{jp}) 
& = & 0\, ,
\\
& & \nonumber \\
\mathcal{D}\mathcal{B}(G_{i}) 
-\epsilon_{ij}\left[\mathcal{B}(H_{m})F^{j}+H_{m}\mathcal{B}(F^{jm})\right]
& = & 0\, ,
\\
& & \nonumber \\
\mathcal{D}\mathcal{B}(\tilde{H}^{m}) 
-\epsilon_{ij}\left[\mathcal{B}(G^{i})F^{jm}+G^{i}\mathcal{B}(F^{jm}) \right]
+\epsilon^{mnp}\mathcal{B}(H_{n})H_{p}
+v^{i}\mathcal{B}(\tilde{F}_{im}) 
& = & 0\, ,
\\
& & \nonumber \\
\mathcal{D}\mathcal{B}(\tilde{F}_{im}) 
+\epsilon_{ij}\epsilon_{mnp}
\left[\mathcal{B}(\tilde{H}^{n})F^{jp}+\tilde{H}^{n}\mathcal{B}(F^{jp})\right]
& & \nonumber \\
& & \nonumber \\
+\epsilon_{ij}\left[\mathcal{B}(G^{j})H_{m}+G^{j}\mathcal{B}(H_{m}) \right]
+v_{i}\mathcal{B}(K_{m})
& = & 0\, ,
\\
& & \nonumber \\
\mathcal{D}\mathcal{B}(K_{m})
+\epsilon_{mnp}\left[\mathcal{B}(F^{in})\tilde{F}_{ip}
+F^{in}\mathcal{B}(\tilde{F}_{ip})\right]
& & \nonumber \\
& & \nonumber \\
-\epsilon_{mnp}\left[\mathcal{B}(\tilde{H}^{n})H_{p}
+\tilde{H}^{n}\mathcal{B}(H_{p}) \right]
& = & 0\, .
\\
& & \nonumber \\
\mathcal{D}\mathcal{B}(K_{a})
+T_{A}{}^{m}{}_{n}\left[\mathcal{B}(F^{in})\tilde{F}_{im}
+F^{in}\mathcal{B}(\tilde{F}_{im})\right]
& & \nonumber \\
& & \nonumber \\
+T_{a}{}^{m}{}_{n}\left[\mathcal{B}(\tilde{H}^{n})H_{m}
+\tilde{H}^{n}\mathcal{B}(H_{m}) \right]
& & \nonumber \\
& & \nonumber \\
-v_{i}T_{a}{}^{m}{}_{n}\mathcal{B}(L_{n}{}^{im}) 
-v_{i}f_{ma}{}^{b}\mathcal{B}(L_{b}{}^{im}) 
+v^{i}T_{a}{}^{m}{}_{n}\mathcal{B}(L_{imnn}) 
& = & 0\, .
\\
& & \nonumber \\
\mathcal{D}\mathcal{B}(K_{\alpha})
+T_{\alpha}{}^{i}{}_{j}\left[\mathcal{B}(F^{jm})\tilde{F}_{im}
+F^{jm}\mathcal{B}(\tilde{F}_{im})\right]
+T_{\alpha\, ij}G^{i}\mathcal{B}(G^{j})
& & \nonumber \\
& & \nonumber \\
+T_{\alpha\, ij}v^{j}\mathcal{B}(L_{m}{}^{im}) 
+T_{\alpha}{}^{i}{}_{j}v^{j}\mathcal{B}(L_{imm}) 
& = & 0\, .
\end{eqnarray}

\subsection{Duality relations}
\label{sec-duality}

\begin{eqnarray}
  \star G^{i}
  &  = &
  \epsilon^{ij}\mathcal{W}_{jk}G^{k}\, ,
  \hspace{1cm} 
(G^{2} =  \tilde{G} \equiv e^{-\varphi} \star G +a G )\, , \\
  & & \nonumber \\
  \tilde{H}^{m}
  & = &
  \mathcal{M}^{mn}\star H_{n}\, ,
  \\
  & & \nonumber \\
  \tilde{F}_{im}
  & = &
  \mathcal{W}_{ij}\mathcal{M}_{mn}\star F^{jn}\, ,
  \\
  & & \nonumber \\
  K_{m}
  & = &
  -\star j_{m}^{(\sigma)}\, ,
  \\
  & & \nonumber \\
  K_{a}
  & = &
  -\star j_{a}^{(\sigma)}\, ,
  \\
  & & \nonumber \\
  K_{\alpha}
  & = &
  -\star j_{\alpha}^{(\sigma)}\, ,
  \\
  & & \nonumber \\
  L_{n}{}^{im}
  & = &
  \star \frac{\partial V}{\partial \vartheta_{im}{}^{n}}\, ,
  \\
  & & \nonumber \\
  L_{a}{}^{im}
  & = &
  \star \frac{\partial V}{\partial \vartheta_{im}{}^{a}}\, ,
  \\
  & & \nonumber \\
  L_{imn}
  & = &
  \star \frac{\partial V}{\partial Z^{imn}}\, .
\end{eqnarray}


\end{document}